\documentclass[11pt,a4paper]{article}
\usepackage[hyperref]{eacl2021}
\usepackage{times}
\usepackage{latexsym}

\usepackage{microtype}

\usepackage{times}
\usepackage{latexsym}

\usepackage{helvet}  
\usepackage{courier}  
\usepackage{url}  
\usepackage{graphicx}  
\usepackage{amsfonts}       
\usepackage{amsmath}
\usepackage{amssymb}
\usepackage{bbm}
\usepackage{amsthm}
\usepackage{pifont}
\usepackage{bm}
\usepackage{booktabs}       
\usepackage{soul}
\usepackage{multirow}
\usepackage{subcaption}

\usepackage[ruled,vlined]{algorithm2e}
\usepackage{makecell}
\aclfinalcopy 
\def\aclpaperid{319} 

\title{TrNews: Heterogeneous User-Interest Transfer Learning for News Recommendation}

\author{Guangneng Hu\\
  HKUST / Hong Kong, China \\
  \texttt{njuhgn@gmail.com} \\\And
  Qiang Yang \\
  HKUST / Hong Kong, China \\
  \texttt{qyang@cse.ust.hk} \\}

\date{}

\begin{document}
\maketitle
\begin{abstract}
We investigate how to solve the cross-corpus news recommendation for unseen users in the future. This is a problem where traditional content-based recommendation techniques often fail. Luckily, in real-world recommendation services, some publisher (e.g., Daily news) may have accumulated a large corpus with lots of consumers which can be used for a newly deployed publisher (e.g., Political news). To take advantage of the existing corpus, we propose a transfer learning model (dubbed as TrNews) for news recommendation to transfer the knowledge from a source corpus to a target corpus. To tackle the heterogeneity of different user interests and of different word distributions across corpora, we design a translator-based transfer-learning strategy to learn a representation mapping between source and target corpora. The learned translator can be used to generate representations for unseen users in the future. We show through experiments on real-world datasets that TrNews is better than various baselines in terms of four metrics. We also show that our translator is effective among existing transfer strategies.
\end{abstract}

\section{Introduction}

News recommendation is key to satisfying users' information need for online services. Some news articles, such as breaking news, are manually selected by publishers and displayed for all users. A huge number of news articles generated everyday make it impossible for editors and users to read through all of them, raising the issue of information overload. Online news platforms provide a service of personalized news recommendation by learning from the past reading history of users, e.g., Google~\cite{das2007google,liu2010personalized}, Yahoo~\cite{trevisiol2014cold,okura2017embedding}, and Bing news~\cite{lu2015content,wang2018dkn}.

When a new user uses the system (cold-start users) or a new article is just created (cold-start items), there are too few observations for them to train a reliable recommender system. Content-based techniques exploit the content information of news (e.g., words and tags) and hence new articles can be recommended to existing users~\cite{pazzani2007content}. Content-based recommendation, however, suffers from the issue of data sparsity since there is no reading history for them to be used to build a profile~\cite{park2009pairwise}.

Transfer learning is a common technique for alleviating the issues of data sparsity~\cite{pan2010transfer,cantador2015cross,liu2018transferable}. A user may have access to many websites such as Twitter.com and Youtube.com~\cite{roy2012socialtransfer,huang2016transferring}, and consume different categories of products such as movies and books~\cite{li2009can}. In this case, transfer learning approaches can recommend articles to a new user in the target domain by exploiting knowledge from the relevant source domains for this new user.

\begin{figure}
\centering
\begin{subfigure}{.49\textwidth}
    \centering
    \includegraphics[height=0.6in,width=7.8cm]{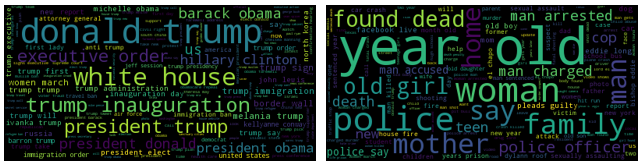}
\end{subfigure}%

\begin{subfigure}{.49\textwidth}
    \centering
    \includegraphics[height=0.6in,width=7.8cm]{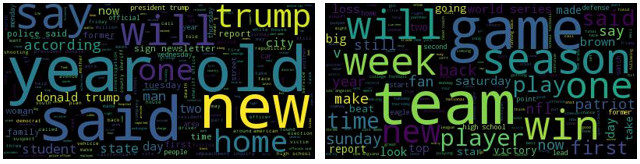}
\end{subfigure}
\caption{Word clouds of two news corpora. Top: Cheetah Mobile data. Bottom: MIND data. Left and right parts represent different domains (categories).}
\label{fig:word-clouds}
\end{figure}

A technical challenge for transfer learning approaches is that user interests are quite different across domains (corpora). For example, users do not use Twitter for the same purpose. A user may follow up on news about ``Donald Trump'' because she supports republican party (in the political news domain), while she may follow up account {\it @taylorswift13} (``Taylor Swift'') because she loves music (in the entertainment news domain). Another challenge is that the word distribution and feature space are different across domains. For example, vocabularies are different for describing political news and entertainment news. An illustration is depicted in Figure~\ref{fig:word-clouds}. As a result, the user profile computed from her news history is heterogeneous across domains.

Several strategies have been proposed for heterogeneous transfer learning~\cite{yang2009heterogeneous}. The transferable contextual bandit (TCB)~\cite{liu2018transferable} learns a translation matrix to translate target feature examples to the source feature space. This linear mapping strategy is also used in collaborative cross networks (CoNet)~\cite{hu2018conet} and deep dual transfer cross domain recommendation (DDTCDR)~\cite{li2020ddtcdr}. To capture complex relations between source and target domains, some nonlinear mapping strategy is considered in the embedding and mapping cross-domain recommendation (EMCDR)~\cite{man2017cross} which learns a supervised regression between source and target factors using a multilayer perceptron (MLP). Since aligned examples between source and target domains are limited, they may face the overfitting issues.

To tackle challenges of heterogeneous user interests and limited aligned data between domains, we propose a novel transfer learning model (TrNews) for cross-corpora news recommendation. TrNews builds a bridge between two base networks (one for each corpus, see Section~\ref{paper:base-network}) through the proposed translator-based transfer strategy. The translator in TrNews captures the relations between source and target domains by learning a nonlinear mapping between them (Section~\ref{paper:translator}). The heterogeneity is alleviated by translating user interests across corpora. TrNews uses the translator to transfer knowledge between source and target networks. TrNews alleviates the limited data in a way of alternating training (Section~\ref{paper:learning}). The learned translator is used to infer the representations of unseen users in the future (Section~\ref{paper:inference}). By ``translating'' the source representation of a user to the target domain, TrNews offers an easy solution to create unseen users' target representations. TrNews outperforms the state-of-the-art recommendation methods on four real-world datasets in terms of four metrics (Section~\ref{exp:recommendation}), while having an explanation advantage by allowing the visualization of the importance of each news article in the history to the future news (Section~\ref{paper:exp-analysis}).

\section{Related Work}

\noindent
{\bf Content recommendation} Content-based recommendation exploits the content information about items (e.g., news title and article body~\cite{yan2012tweet,xiao2019beyond,ma2019news2vec,wu-etal-2020-mind,hu-etal-2020-graph}, tag, vlog~\cite{gao2010vlogging}), builds a profile for each user, and then matches users to items~\cite{lops2011content,yu2016user,wu2019neural-emnlp}. It is effective for items with content or auxiliary information but suffers from the issues of data sparsity for users. DCT~\cite{barjasteh2015cold} constructs a user-user similarity matrix from user demographic features including gender, age, occupation, and location~\cite{park2009pairwise}. NT-MF~\cite{huang2016transferring} constructs a user-user similarity matrix from Twitter texts. BrowseGraph~\cite{trevisiol2014cold} addresses the freshly news recommendation by constructing a graph using URL links between web pages. NAC~\cite{rafailidis2019neural} transfers from multiple source domains through the attention mechanism. PdMS~\cite{felicio2017multi} assumes that there are many recommender models available to select items for a user, and introduces a multi-armed bandit for model selection. LLAE~\cite{li2019zero} needs a social network as side information for cold-start users. Different from the aforementioned works, we aim to recommending news to unseen users by transferring knowledge from a source domain to a target domain.

\begin{figure*}
\begin{subfigure}{.69\textwidth}
  \includegraphics[height=1.8in,width=10.2cm]{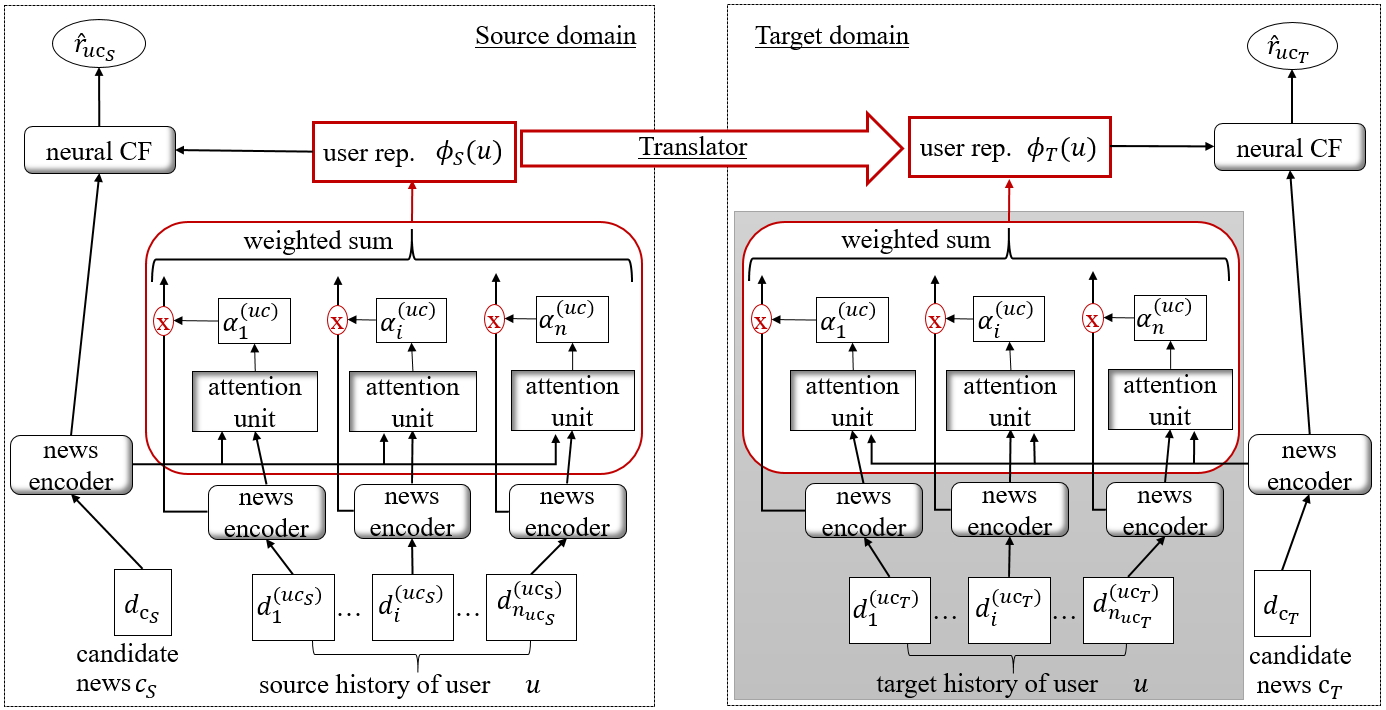}
  \caption{TrNews }
  \label{fig:trnews}
\end{subfigure}%
\begin{subfigure}{.25\textwidth}
  \includegraphics[height=1.1in,width=5.cm]{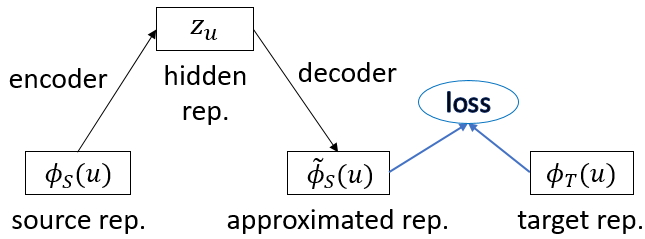}
  \caption{Translator }
  \label{fig:translator}
\end{subfigure}
\caption{Left: Architecture of TrNews. There is a base network for each of the two domains. The shaded area in the target network is empty for unseen users. The translator enables knowledge transfer between source and target networks. Right: The translator.}
\centering
\label{fig:trnews-translator}
\end{figure*}

\noindent
{\bf Transfer learning} Transfer learning aims at improving the performance of a target domain by exploiting knowledge from source domains~\cite{pan2009survey}. A special setting is domain adaptation where a source domain provides labeled training examples while the target domain provides instances on which the model is meant to be deployed~\cite{glorot2011domain,li2019semi}. The coordinate system transfer (CST)~\cite{pan2010transfer} firstly learns the principle coordinate of users in the source domain, and then transfers it to the target domain in the way of warm-start initialization. This is equivalent to an identity mapping from users' source representations to their corresponding target representations. TCB~\cite{liu2018transferable} learns a linear mapping to translate target feature examples to the source feature space because there are many labelled data in the source domain. This linear strategy is also used in CoNet~\cite{hu2018conet} and DDTCDR~\cite{li2020ddtcdr} which transforms the source representations to the target domain by a translation matrix. Nonlinear mapping strategy~\cite{man2017cross,zhu2018deep,fu2019deeply} is to learn a supervised mapping function between source and target latent factors by using neural networks. SSCDR~\cite{kang2019semi} extends them to the semi-supervised mapping setting. Our translator is general to accommodate these identity, linear, and nonlinear transfer-learning strategies.

\section{TrNews}\label{paper:trnews}


\subsection{Architecture}\label{paper:architecture}

The architecture of TrNews is shown in Figure~\ref{fig:trnews}, which has three parts. There are a source network for the source domain $S$ and a target network for the target domain $T$, respectively. The source and target networks are both an instantiation of the base network (Section~\ref{paper:base-network}). The translator enables knowledge transfer between the two networks (Section~\ref{paper:translator}). We give an overview of TrNews before introducing the base network and the translator.

\noindent
{\bf Target network} The information flow goes from the input, i.e., (user $u$, candidate news $c_T$) to the output, i.e., the preference score $\hat{r}_{uc_T}$, through the following three steps. First, the news encoder $\psi_T$ computes the news representation from its content. The candidate news representation is $\psi_T(c_T)=\psi_T(d_{c_T})$ where $d_{c_T}$ is $c_T$'s content. The representations of historical news articles $[i]_{i=1}^{n_{uc_T}}$ of the user are $[\psi_T(d_i^{(uc_T)})]_{i=1}^{n_{uc_T}}$ where $d_i^{(uc_T)}$ is $i$'s content and $n_{uc_T}$ is size of the history. Second, the user encoder $\phi_T$ computes the user representation from her news history by: $\phi_T(u) = \phi_T\Big([\psi_T(d_i^{(uc_T)})]_{i=1}^{n_{uc_T}}\Big)$. Third, the neural collaborative filtering (CF) module $f_T$ computes the preference score by: $\hat{r}_{uc_T} = f_T([\phi_T(u),\psi_T(c_T)])$. We can denote the target network by a tuple $(\psi_T, \phi_T, f_T)$.

\noindent
{\bf Source network} Similarly to the three-step computing process in target network, we compute preference score $\hat{r}_{uc_S}$ from input ($u, c_S$) by: $\hat{r}_{uc_S} = f_S([\phi_S(u),\psi_S(c_S)])$ with tuple $(\psi_S, \phi_S, f_S)$.

\noindent
{\bf Translator} The translator $\mathcal{F}$ learns a mapping from the user's source representation to her target representation by $\mathcal{F}: \phi_S(u) \rightarrow \phi_T(u)$.

\subsubsection{Base network}\label{paper:base-network}

There is a base network for each of the two domains. It is an attentional network which has three modules $(\psi, \phi, f)$: the news encoder $\psi$ to learn news representations, the user encoder $\phi$ to learn user representations, and a neural collaborative filtering module $f$ to learn user preferences from reading behaviors.

\noindent
{\bf News encoder} The news encoder module is to learn news representation from its content. The news encoder takes a news article $c$'s word sequence $d_c = [w_j]_{j=1}^{n_c}$ ($n_c$ is length of $c$) as the input, and outputs its representation $\psi(c) \triangleq \psi(d_c) \in \mathbb{R}^D$ where $D$ is the dimensionality. We compute the average of $c$'s word embeddings by: $\psi(d_c) = \frac{1}{|d_c|} \sum_{w \in d_c} e_w,$ where $e_w$ is the embedding of $w$.

\noindent
{\bf User encoder} The user encoder module is to learn the user representation from their reading history. The user encoder takes a user's reading history $[\psi(d_i^{(u)})]_{i=1}^{n_u}$ ($n_u$ is length of $u$'s history) as input, and outputs her representation $\phi(u) \triangleq \phi([\psi(d_i^{(u)})]_{i=1}^{n_u}) \in \mathbb{R}^D$ where $D$ is dimensionality.

In detail, given a pair of user and candidate news $(u,c)$, we get the user representation $\phi(u|c)$ as the weighted sum of her historical news articles' representations: $\phi(u|c) = \sum_{i=1}^{n_{uc}} \alpha_i^{(uc)} \psi(d_i^{(u)})$. The weights $\alpha^{(uc)}_i$'s are computed via attention units by: $\alpha_i^{(uc)} = a([\psi(d_i^{(u)}),\psi(d_c)])$ where $a$ is the attention function with parameters to be learned. We use an MLP to compute it. For a specific candidate news $c$, we limit the history news to only those articles that are read before it. For notational simplicity, we do not explicitly specify the candidate news when referring to a user representation, i.e., $\phi(u)$ for short of $\phi(u|c)$.

\noindent
{\bf Neural CF} The neural collaborative filtering module is to learn preferences from user-news interactions. The module takes concatenated representations of user and news $[\phi(u),\psi(c)]$ as input, and outputs preference score $\hat{r}_{uc} = f([\phi(u),\psi(c)])$ where $f$ is an MLP.

\subsection{Translator}\label{paper:translator}

The target network suffers from the data sparsity issue of users who have no reading history. In this section, we propose a transfer learning component (i.e., the translator) to enable knowledge transfer for cross-domain news recommendation. The challenge is that user interests and word distributions are different across domains. For example, we compute the word clouds for two news corpora as shown in Figure~\ref{fig:word-clouds}. We can see that their word distributions are quite different and vocabularies are also different. Hence, user representations computed from their news history are heterogeneous across domains.

We build a translator, $\mathcal{F}: \phi_S(u) \rightarrow \phi_T(u)$, to learn a mapping from a user's source representation to her target representation as shown in Figure~\ref{fig:translator}. This translator captures the relationship and heterogeneity across domains. The translator learns to approximate the target representation from the source representation.

The translator takes a user's source representation $\phi_S(u)$ as the input, and maps it to a hidden representation $z_u$ via an encoder parameterized by $\theta$, and then gets a approximated representation ${\tilde\phi_S(u)}$ from it via a decoder parameterized by $\theta'$. The parameters $\Theta_\mathcal{F} = \{\theta, \theta'\}$ of the translator are optimized to minimize the approximation error:
\begin{equation}
\scalebox{0.8}[1]{$\mathcal{L}_\mathcal{F} = \frac{1}{|\mathcal{U}_0|} \sum\nolimits_{u \in \mathcal{U}_0} ||{H \tilde\phi_S(u)} - \phi_T(u)||_2^2$},
\label{eq:translator}
\end{equation}
where $\mathcal{U}_0 = \mathcal{U}_S \cap \mathcal{U}_T$, and $\mathcal{U}_S$ and $\mathcal{U}_T$ are the user sets of source and target domains, respectively. $H$ is to match the dimensions of source and target representations.

Note that, we do not minimize the approximation error between $\phi_S(u)$ and ${\tilde\phi_S(u)}$ as with the standard autoencoder because our goal is to learn a mapping from a user's source representation to her corresponding target representation. After training, the learned mapping function is then used for inferring representations of unseen users in the target domain (the inference process will be described later in Section~\ref{paper:inference}). It fulfills knowledge transfer from the source to the target domain via a supervised learning process.

\noindent
{\bf Extensions} The translator can be generalized to multiple, say $k$, source domains. We learn $k$ translators using the aligned examples from each of the source domain to the target domain and then we average (or concatenate) the $k$ mapped representations as the final representation for the user. Another extension is to introduce denoising or stacking techniques into the translator framework, not just the MLP structure in~\cite{man2017cross}.

\subsection{Model learning}\label{paper:learning}

We learn TrNews in two stages. First, we train the source network using source training examples $D_S$ and train the target network using target training examples $D_T$, respectively. Second, we train the translator by pairs of user representations computed on-the-fly from source and target networks. We introduce these two stages in detail.

First, TrNews optimizes the parameters associated with target network $\Theta_T = \{\theta_{\phi_T}, \theta_{\psi_T}, \theta_{f_T}\}$ and source network $\Theta_S = \{\theta_{\phi_S}, \theta_{\psi_S}, \theta_{f_S}\}$ by minimizing the joint cross-entropy loss:
\begin{multline}\label{eq:loss}
\scalebox{0.8}[1]{$\mathcal{L} = - \sum_{D_T} (r_{uc_T}\log\hat{r}_{uc_T} + (1-r_{uc_T})\log(1-\hat{r}_{uc_T}))$} \\ \scalebox{0.8}[1]{$- \sum_{ D_S} (r_{uc_S}\log\hat{r}_{uc_S} + (1-r_{uc_S})\log(1-\hat{r}_{uc_S}))$},
\end{multline}
where the two terms on the right-hand side are to optimize losses over user-news examples in the target and source domains, respectively. They are related by the word embedding matrix for the union of words of the two domains. We generate $D_T$ and $D_S$ as follows and take the target domain as an example since the procedure is the same for the source domain. Suppose we have a whole news reading history for a user $u$, say $[d_1,d_2,...,d_{n_u}]$. Then we generate the positive training examples by sliding over the history sequence: $D_T^+ = \{([d_i]_{i=1}^{c-1}, d_c): c=2,...,n_u\}$. We adopt the random negative sampling technique~\cite{pan2008one} to generate the corresponding negative training examples $D_T^- = \{([d_i]_{i=1}^{c-1}, d_c'): d_c' \notin [d_1,d_2,...,d_{n_u}]\}$, that is, we randomly sample a news article from the corpus as a negative sample which is not in this user's reading history.

Second, TrNews optimizes the parameters associated with the translator $\Theta_\mathcal{F} = \{\theta, \theta'\}$ by Eq.~(\ref{eq:translator}). Since the aligned data is limited, we increase the training pairs by generating them on-the-fly during the training of the two networks, i.e., in an alternating way. The model learning is summarized in Algorithm~\ref{algo:training}.

\begin{algorithm}\small
\SetAlgoLined
{\bf Input}: $D_T, D_S, \mathcal{U}_0$  \\
{\bf Output}: Source \& target networks,  translator \\
 \For{$iter=1,2,...,50$}{
 \begin{enumerate}
 \item Train target and source networks with  \\
       mini batch using $D_T, D_S$ respectively\;
 \item \For{$u \in \mathcal{U}_0$}{
        \begin{enumerate}
         \item Generate source representations \\
               $\phi(u_S)$ using source network\;
         \item Generate target representations \\
               $\phi(u_T)$ using target network\;
         \item Train the translator using pairs \\ $(\phi(u_S), \phi(u_T))$ with mini batch\;
         \end{enumerate}
        }
 \item Compute metrics on the validation set\;
     \If{No improvement for 10 iters}{
       $\quad \quad $ Early stopping\;
      }
 \end{enumerate}
 }
 \caption{Training of TrNews.}
 \label{algo:training}
\end{algorithm}

\subsection{Inference for unseen users}\label{paper:inference}

For a new user in the target domain (not seen in the training set $\mathcal{U}_T^{train}$), we do not have any previous history to rely on in learning a user representation for her. That is, the shaded area of the target network in Figure~\ref{fig:trnews} is empty for unseen users.

TrNews estimates a new user $u^*$'s target representation by mapping from her source representation using the learned translator $\mathcal{F}$ by:
\begin{equation}
\scalebox{0.8}[1]{$\phi_T(u^*) := \mathcal{F}(\phi_S(u^*)), \; \forall u^* \in \mathcal{U}_S \land u^* \notin \mathcal{U}_T^{train}$},
\end{equation}
where we compute $\phi_S(u^*)$ using $u$'s latest reading history in the source domain. Then we can predict the user preference for candidate news $c^*$ by:
\begin{equation}
\scalebox{0.8}[1]{$\hat{r}_{u^* c^*} = f_T([\phi_T(u^*),\psi_T(c^*)])$}.
\end{equation}


\section{Experiment}\label{paper:exp}

We evaluate the performance of TrNews (Section~\ref{exp:recommendation}) and the effectiveness of the translator (Section~\ref{exp:translator}) in this section.

\subsection{Datasets and experimental setup}

\noindent
{\bf Datasets} We evaluate on two real-world datasets. The first {\it {\bm{$NY, FL, TX , \& CA$}}} are four subdatasets extracted from a large dataset provided by an internet company Cheetah Mobile~\cite{liu2018transferable,hu2019transfer}. The information contains news reading logs of users in a large geographical area collected in January of 2017, ranging from New York (NY), Florida (FL), Texas (TX), to California (CA) based on the division of user geolocation. They are treated as four rather than a single because the user set is not overlapped among them. The top two categories (political and daily) of news are used as the cross corpora. The mean length of news articles is around 12 words while the max length is around 50 words. The mean length of user history is around 45 articles while the max length is around 900 articles.  The second {\it {\bm{$MIND$}}} is a benchmark dataset released by Microsoft for news recommendation~\cite{wu-etal-2020-mind}. We use the MIND-small version to investigate the knowledge transfer when news reading examples are not so large and it is publicly available~\url{https://msnews.github.io/}. The title and abstract of news are used as the content. The clicked historical news articles are the positive examples for user. The top two categories (news and sports) of news are used as the cross corpora. The word clouds of the two datasets are shown in Figure~\ref{fig:word-clouds} and the statistics are summarized in Table~\ref{tb:data}. The mean length of news articles is around 40 words while the max length is around 123 words. Besides, the mean length of user history is around 13 articles while the max length is around 246 articles.


\noindent
{\bf Evaluation protocol} We randomly split the whole user set into two parts, training and test sets where the ratio is 9:1. Given a user in the test set, for each news in her history, we follow the strategy in~\cite{he2017neural} to randomly sample a number of negative news, say 99, which are not in her reading history and then evaluate how well the recommender can rank this positive news against these negative ones. For each user in the training set, we reserve her last reading news as the valid set. We follow the typical metrics to evaluate top-$K$ news recommendation~\cite{peng2016news,okura2017embedding,an2019neural} which are hit ratio (HR), normalized discounted cumulative gain (NDCG), mean reciprocal rank (MRR), and the area under the ROC curve (AUC). We report the results at cut-off $K \in \{5,10\}$.

\begin{table}\small			
\centering	
\resizebox{0.49\textwidth}{!}{					
\begin{tabular}{|c | c | c|c|c | c|c|c|}		
\hline 								
{\multirow{2}{*}{Data}} & \multirow{2}{*}{\#user} & \multicolumn{3}{c|}{Target domain}  & \multicolumn{3}{|c|}{Source domain}\\						
\cline{3-5}	\cline{6-8}				 		
\multicolumn{1}{|c|}{}  &   &   \#news &  \#reading  &  \#word  & \#news & \#reading & \#word   \\		
\hline
NY &    14,419    & 33,314 & 158,516 & 368,000 & 23,241 & 139,344 & 273,894  \\
\hline
FL  &    15,925  & 33,801  & 178,307 & 376,695 & 25,644 & 168,081 & 340,797   \\
\hline
TX  &    20,786  & 38,395  & 218,376 & 421,586 & 29,797 & 221,344 & 343,706  \\
\hline
CA  &    26,981  & 44,143  & 281,035 & 481,959 & 32,857 & 258,890 & 375,612   \\
\hline
MIND&    25,580  & 9,372   & 211,304 & 461,984 & 8,577  & 120,409 & 346,988   \\
\hline
\end{tabular}	
}	
\caption{Statistics of the datasets. }
\label{tb:data}										
\end{table}

\noindent
{\bf Implementation} We use TensorFlow. The optimizer is Adam~\cite{kingma2015adam} with learning rate 0.001. The size of mini batch is 256. The neural CF module has two hidden layers with size 80 and 40 respectively. The size of word embedding is 128. The translator has one hidden layer on the smaller datasets and two on the larger ones. The history is the latest 10 news articles.


\begin{table}
\centering
\resizebox{.5\textwidth}{!}{
\begin{tabular}{|c | l| l| l| l| l| l|}
\hline
\bf{NY}  & HR@5    & HR@10   & NDCG@5  & NDCG@10 & MRR     & AUC \\
\hline
POP      & 52.96 & 67.66 & 40.34* & 45.10 & 39.89* & 77.92 \\
\hline
LR       & 53.24 & 74.00 & 36.15 & 42.86 & 34.95 & 91.64 \\
\hline
TANR     & 52.53 & 71.63 & 37.24 & 43.37 & 36.50 & 91.35 \\
\hline
DeepFM   & 52.02 & 73.71 & 39.17 & 45.38 & 39.56 & 91.79 \\
\hline
DIN      & 57.10* & 75.66* & 40.23 & 46.13* & 38.65 & 92.29* \\
\hline
TrNews   & {\bf 82.60} & {\bf 95.15} & {\bf 60.78} & {\bf 64.83} & {\bf 55.70} & {\bf 97.28} \\
\hline
\end{tabular}
}

\resizebox{.5\textwidth}{!}{
\begin{tabular}{|c | l| l| l| l| l| l|}
\hline
\bf{FL}   & HR@5    & HR@10   & NDCG@5  & NDCG@10 & MRR     & AUC \\
\hline
POP      & 52.45 & 66.14 & 39.72* & 44.15 & 39.15* & 79.33 \\
\hline
LR       & 54.26* & 73.90* & 37.15 & 43.56 & 35.89 & 91.79 \\
\hline
TANR     & 49.98 & 69.46 & 36.08 & 42.37 & 35.95 & 90.88 \\
\hline
DeepFM   & 52.36 & 73.02 & 36.05 & 42.74 & 36.29 & 91.64 \\
\hline
DIN      & 53.98 & 73.33 & 37.96 & 44.18* & 36.96 & 91.86* \\
\hline
TrNews   & {\bf 81.83} & {\bf 94.45} & {\bf 62.53} & {\bf 66.63} & {\bf 58.39} & {\bf 97.41} \\
\hline
\end{tabular}
}

\resizebox{.5\textwidth}{!}{
\begin{tabular}{|c | l| l| l| l| l| l|}
\hline
\bf{TX}   & HR@5  & HR@10   & NDCG@5  & NDCG@10 & MRR     & AUC \\
\hline
POP      & 54.21 & 67.87 & 40.62* & 45.03* & 39.64* & 81.31 \\
\hline
LR       & 55.72* & 73.80* & 39.24 & 44.97 & 37.78 & 91.74* \\
\hline
TANR     & 49.87 & 68.75 & 35.82 & 41.89 & 35.59 & 90.56 \\
\hline
DeepFM   & 52.19 & 71.95 & 35.40 & 41.92 & 35.65 & 91.17 \\
\hline
DIN      & 53.72 & 72.70 & 38.47 & 44.59 & 37.62 & 91.53 \\
\hline
TrNews   & {\bf 81.50} & {\bf 94.67} & {\bf 61.76} & {\bf 66.11} & {\bf 57.49} & {\bf 97.21} \\
\hline
\end{tabular}
}

\resizebox{.5\textwidth}{!}{
\begin{tabular}{|c | l| l| l| l| l| l|}
\hline
\bf{CA}   & HR@5    & HR@10   & NDCG@5  & NDCG@10 & MRR     & AUC \\
\hline
POP      & 58.32* & 71.19 & 44.71* & 48.86* & 43.44* & 83.38 \\
\hline
LR       & 58.82 & 75.67* & 42.16 & 47.65 & 40.44 & 92.37*  \\
\hline
TANR     & 49.87 & 68.75 & 35.81 & 41.88 & 35.58 & 90.56 \\
\hline
DeepFM   & 55.58 & 74.73 & 38.82 & 45.16 & 38.21 & 92.25 \\
\hline
DIN      & 55.31 & 73.70 & 40.14 & 46.09 & 39.20 & 92.03  \\
\hline
TrNews   & {\bf 81.54} & {\bf 94.72} & {\bf 61.99} & {\bf 66.25} & {\bf 57.70} & {\bf 97.22} \\
\hline
\end{tabular}
}

\resizebox{.5\textwidth}{!}{
\begin{tabular}{|c | l| l| l| l| l| l|}
\hline
\bf{MIND}   & HR@5    & HR@10   & NDCG@5  & NDCG@10 & MRR     & AUC \\
\hline
POP      & 84.18  & 92.80  & 69.61 & 72.43 & 66.33  & 95.13 \\
\hline
LR       & 92.69*  & 96.66*  & 85.81 & 87.11* & 84.33*  & 97.92* \\
\hline
TANR     & 89.94  & 95.34  & 89.94* & 83.38 & 79.84  & 97.86  \\
\hline
DeepFM   & 89.16  & 94.78  & 79.36 & 81.19 & 77.12  & 97.63 \\
\hline
DIN      & 89.28  & 94.88  & 80.16 & 82.03 & 78.22  & 97.63  \\
\hline
TrNews   & {\bf 97.36}  & {\bf 99.02}  & {\bf 94.16} & {\bf 94.74} & {\bf 93.45}  & {\bf 99.47} \\
\hline
\end{tabular}
}

\caption{Comparison of different recommenders. }
\label{tb:results}
\end{table}

\subsection{Comparing different recommenders}\label{exp:recommendation}

In this section, we show the recommendation results by comparing TrNews with different state-of-the-art methods.

\noindent
{\bf Baselines} We compare with following recommendation methods which are trained on the merged source and target datasets by aligning with shared users: {\bf POP}~\cite{park2009pairwise} recommends the most popular news. {\bf LR}~\cite{mcmahan2013ad} is widely used in ads and recommendation. The input is the concatenation of candidate news and user's representations. {\bf DeepFM}~\cite{guo2017deepfm} is a deep neural network for ads and recommendation based on the wide \& deep structure. We use second-order feature interactions of reading history and candidate news, and the input of deep component is the same as LR. {\bf DIN}~\cite{zhou2018deep} is a deep interest network for ads and recommendation based on the attention mechanism. We use the news content for news representations. {\bf TANR}~\cite{wu2019neural} is a state-of-the-art deep news recommendation model using an attention network to learn the user representation. We adopt the news encoder and negative sampling the same with TrNews.



\noindent
{\bf Results} We have observations from results of different recommendation methods as shown in Table~\ref{tb:results}. Firstly, considering that breaking and headline news articles are usually read by every user, the POP method gets competitive performance in terms of NDCG and MRR since it ranks the popular news higher than the other news. Secondly, the neural methods are generally better than the traditional, shallow LR method in terms of NDCG, MRR, and AUC on the four subdatasets. It may be that neural networks can learn nonlinear, complex relations between the user and the candidate news to capture user interests and news semantics. Considering that the neural representations of user and candidate news are fed as the input of LR, it gets competitive performance on MIND data. Finally, the proposed TrNews model achieves the best performance with a large margin improvement over all other baselines in terms of HR, NDCG, and MRR and also with an improvement in terms of AUC. It validates the necessity of accounting for the heterogeneity of user interests and word distributions across domains. This also shows that the base network is an effective architecture for news recommendation and the translator is effective to enable the knowledge transfer from the source domain to the target domain. In more detail, it is inferior by training a global model from the mixed source and target examples and then using this global model to predict user preferences on the target domain, as baselines do. Instead, it is good by training source and target networks on the source and target domains, respectively, and then learning a mapping between them, as TrNews does.

\begin{table}\small
\centering
\resizebox{.49\textwidth}{!}{
\begin{tabular}{|c|c|c|}
\hline
Approach   & Transfer strategy & Formulation \\
\hline
CST~\cite{pan2010transfer}  & Identity mapping & $\phi_T(u) = \phi_S(u)$ \\
\hline
\begin{tabular}[c]{@{}c@{}}TCB~\cite{liu2018transferable} \\ \cline{1-1}
DDTCDR~\cite{li2020ddtcdr} \end{tabular}  & Linear mapping & \begin{tabular}[c]{@{}c@{}}$\phi_T(u) = H \phi_S(u)$ \\ \cline{1-1}
$H$ is orthogonal \end{tabular}\\
\hline
EMCDR~\cite{man2017cross} & Nonlinear mapping & $\phi_T(u) = \textrm{MLP}(\phi_S(u))$ \\
\hline
\end{tabular}
}
\caption{Different transfer learning strategies.}
\label{tb:tranfser-learning-strategy}
\end{table}

\subsection{Comparing different transfer strategies}\label{exp:translator}

In this section, we demonstrate the effectiveness of the translator-based transfer-learning strategy.

\noindent
{\bf Baselines} We replace the translator of TrNews with the transfer-learning strategies of baseline methods as summarized in Table~\ref{tb:tranfser-learning-strategy}. All baselines are state-of-the-art recommenders and capable of recommending news to cold-start users. Note that, the compared transfer-learning methods are upgraded from their original versions. We strengthen them by using the neural attention architecture as the base component. In their original versions, CST and TCB use matrix factorization (MF) while DDTCDR and EMCDR use multilayer perceptron. The neural attention architecture has shown superior performance over MF and MLP in the literature~\cite{zhou2018deep,wu2019neural}. As a result, we believe that the improvement will be larger if we compare with their original versions but this is obviously unfair.

\begin{table}
\centering
\resizebox{.48\textwidth}{!}{
\begin{tabular}{|c | l| l| l| l| l| l|}
\hline
\bf{NY}      & HR@5    & HR@10   & NDCG@5  & NDCG@10 & MRR     & AUC \\
\hline
CST        & 81.04 & 94.37 & 59.04 & 63.56 & 54.19 & 96.94 \\
\hline
TCB        & 82.18 & 94.92* & 60.36* & 64.46* & 55.23* & 97.28* \\
\hline
DDTCDR     & 82.27 & 94.90 & 59.82 & 63.90 & 54.51 & 97.25 \\
\hline
EMCDR      & 82.44* & 94.87 & 60.35 & 64.33 & 55.06 & 97.24 \\
\hline
TrNews     & {\bf 82.60} & {\bf 95.15} & {\bf 60.78} & {\bf 64.83} & {\bf 55.70} & {\bf 97.28} \\
\hline
\end{tabular}
}

\resizebox{.48\textwidth}{!}{
\begin{tabular}{|c | l| l| l| l| l| l|}
\hline
\bf{FL}     & HR@5    & HR@10   & NDCG@5  & NDCG@10 & MRR     & AUC \\
\hline
CST        & 79.29 & 93.91 & 59.03 & 63.60 & 54.74 & 97.07 \\
\hline
TCB        & 81.51 & {\bf 94.83} & 62.06 & 66.33* & 57.90* & 97.40* \\
\hline
DDTCDR     & 81.39 & 94.63* & 61.76 & 66.12 & 57.68 & 97.37 \\
\hline
EMCDR      & 81.52* & 94.47 & 62.14* & 66.23 & 57.87 & 97.37 \\
\hline
TrNews     & {\bf 81.83} & 94.45 & {\bf 62.53} & {\bf 66.63} & {\bf 58.39} & {\bf 97.41} \\
\hline
\end{tabular}
}

\resizebox{.48\textwidth}{!}{
\begin{tabular}{|c | l| l| l| l| l| l|}
\hline
\bf{TX}     & HR@5    & HR@10   & NDCG@5  & NDCG@10 & MRR     & AUC \\
\hline
CST        & 78.74 & 94.20 & 58.53 & 63.48 & 54.56 & 96.92 \\
\hline
TCB        & 80.68 & 94.12 & 61.06 & 65.38 & 56.97 & 97.10 \\
\hline
DDTCDR     & 81.08 & 94.57 & 61.02 & 65.50 & 56.87 & 97.10 \\
\hline
EMCDR      & 81.34* & {\bf 94.72} & {\bf 61.78} & 66.11* & {\bf 57.59} & 97.16* \\
\hline
TrNews     & {\bf 81.50} & 94.67* & 61.76* & {\bf 66.11} & 57.49* & {\bf 97.21} \\
\hline
\end{tabular}
}

\resizebox{.48\textwidth}{!}{
\begin{tabular}{|c | l| l| l| l| l| l|}
\hline
\bf{CA}     & HR@5    & HR@10   & NDCG@5  & NDCG@10 & MRR     & AUC \\
\hline
CST        & 79.92 & 93.71* & 60.19 & 64.63 & 55.97 & 97.12 \\
\hline
TCB        & 80.90* & 93.71* & {\bf 62.32} & {\bf 66.45} & {\bf 58.35} & {\bf 97.36} \\
\hline
DDTCDR     & 80.22 & 93.47 & 61.42 & 65.72 & 57.44 & 97.25 \\
\hline
EMCDR      & 80.53 & 93.33 & 62.04* & 66.18 & 58.11* & 97.30* \\
\hline
TrNews     & {\bf 81.54} & {\bf 94.72} & 61.99 & 66.25* & 57.70 & 97.22 \\
\hline
\end{tabular}
}

\resizebox{.48\textwidth}{!}{
\begin{tabular}{|c | l| l| l| l| l| l|}
\hline
\bf{MIND}  & HR@5    & HR@10   & NDCG@5  & NDCG@10 & MRR     & AUC \\
\hline
CST     & 96.93  & 98.49  & 93.83  & 94.34  & 93.09 & 99.32 \\
\hline
TCB     & 97.41*  & 98.94  & 94.21*  & 94.72  & 93.43 &  99.41 \\
\hline
DDTCDR  & 97.38  & 98.99  & {\bf 94.25}  &  {\bf 94.78} & {\bf 93.49} &  99.46* \\
\hline
EMCDR   & {\bf 97.42}  & 99.01*  & 94.16  & 94.68  & 93.35 & 99.45 \\
\hline
TrNews  & 97.36  & {\bf 99.02}  & 94.16 & 94.74* & 93.45*  & {\bf 99.47} \\
\hline
\end{tabular}
}

\caption{Comparison of different transfer strategies. }
\label{tb:transfer-learning-results}
\end{table}

\noindent
{\bf Results} We have observations from results of different transfer learning strategies as shown in Table~\ref{tb:transfer-learning-results}. Firstly, the transfer strategy of identity mapping (CST) is generally inferior to the linear (TCB and DDTCDR) and nonlinear (EMCDR and TrNews) strategies. CST directly transfers the source knowledge to the target domain without adaptation and hence suffers from the heterogeneity of user interests and word distributions across domains. Secondly, the nonlinear transfer strategy of EMCDR is inferior to the linear strategy of TCB in terms of MRR and AUC on the two smaller NY and FL datasets. This is probably because EMCDR increases the model complexity by introducing two large fully-connected layers in its MLP component. In contrast, our translator is based on the small-waist  autoencoder-like architecture and hence can resist overfitting to some extent. Finally, our translator achieves the best performance in terms of NDCG, MRR and AUC on the two smaller NY and FL datasets, and achieves competitive performance on the two larger TX and CA datasets, and achieves the best performance in terms of HR and AUC on the MIND dataset, comparing with other four transfer methods. These results validate that our translator is a general and effective transfer-learning strategy to capture the diverse user interests accurately during the knowledge transfer for the unseen users in cross-domain news recommendation.

\subsection{Analysis}\label{paper:exp-analysis}

\begin{figure}
\begin{subfigure}{.235\textwidth}
  \centering
  \includegraphics[height=1.2in,width=3.7cm]{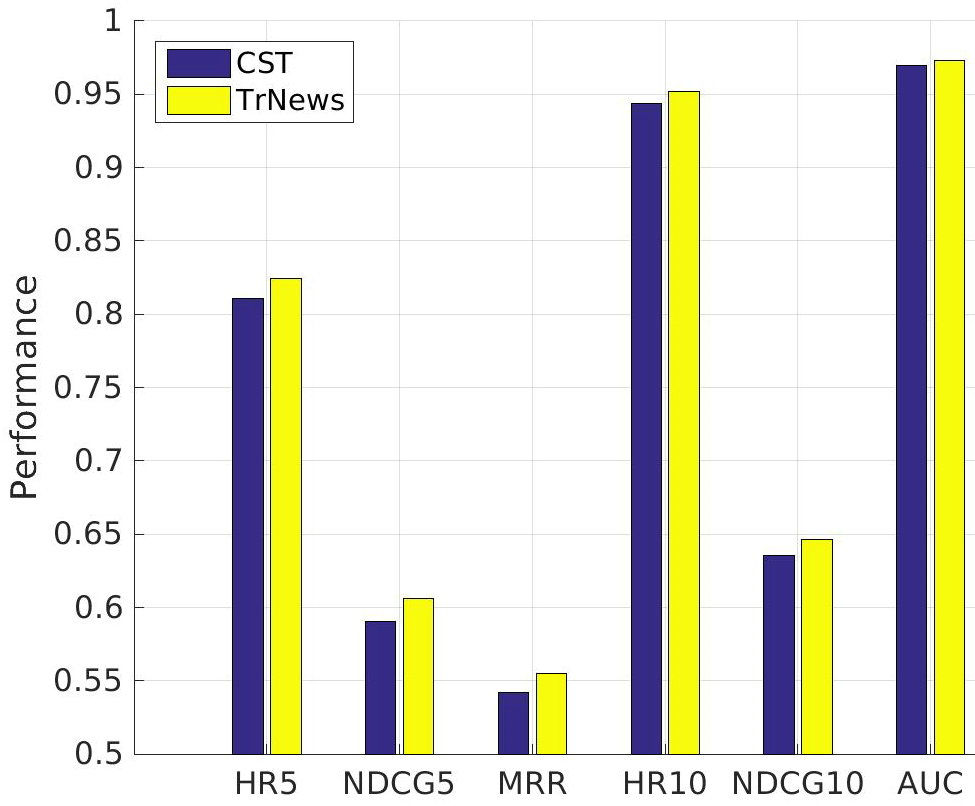}
\end{subfigure}
\begin{subfigure}{.235\textwidth}
  \centering
  \includegraphics[height=1.2in,width=3.7cm]{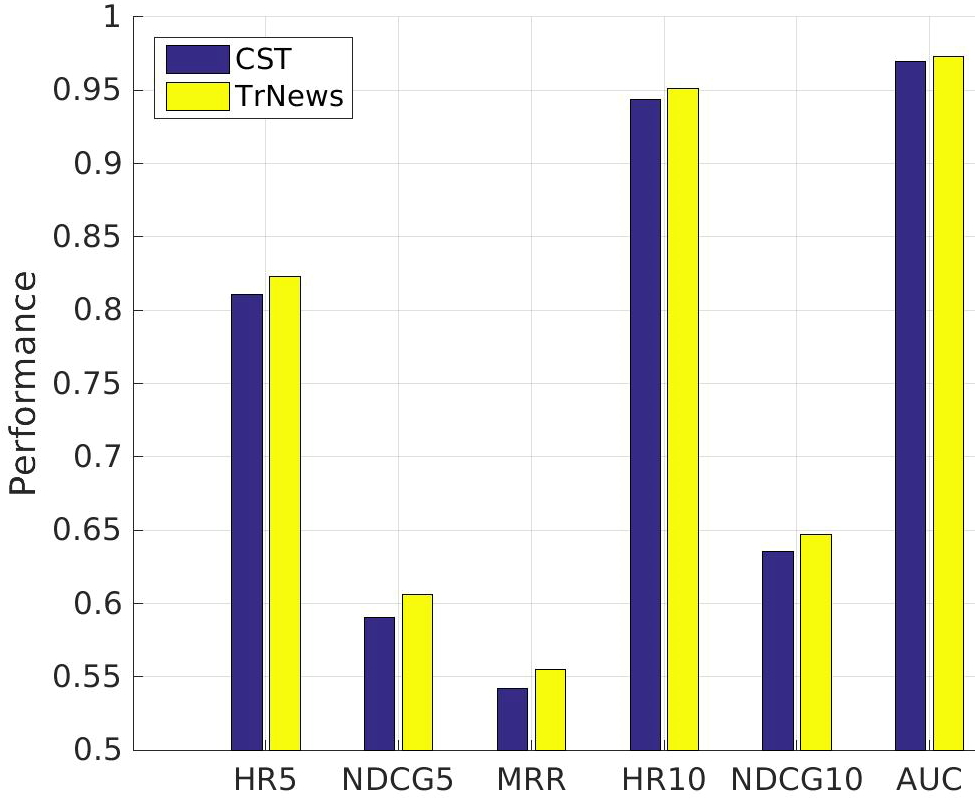}
\end{subfigure}

\begin{subfigure}{.235\textwidth}
  \centering
  \includegraphics[height=1.2in,width=3.7cm]{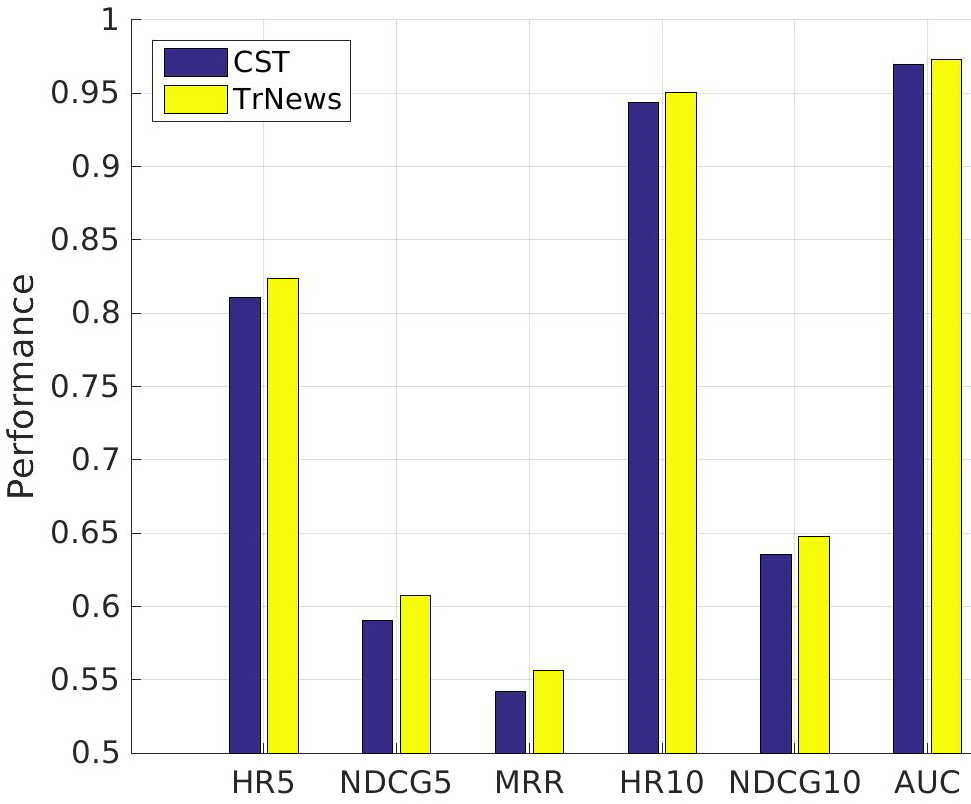}
\end{subfigure}
\begin{subfigure}{.235\textwidth}
  \centering
  \includegraphics[height=1.2in,width=3.7cm]{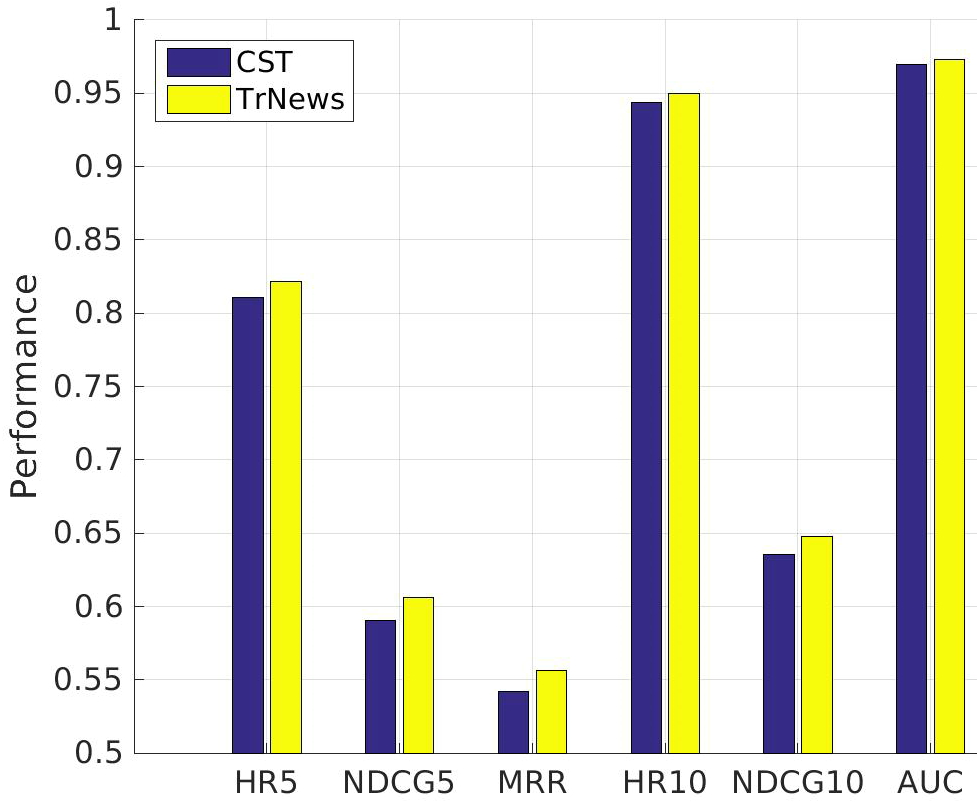}
\end{subfigure}
\caption{Impact of percentage (90\%, 70\%, 50\%, 30\%.) of shared users used to train the translator. }
\label{fig:fraction}
\end{figure}

\noindent
{\bf Benefit of knowledge transfer} We vary the percentage of shared users used to train the translator (see Eq.~(\ref{eq:translator})) with $\{90\%, 70\%, 50\%, 30\%\}$. We compare with a naive transfer strategy of CST, i.e., the way of direct transfer without adaptation. The results are shown in Figure~\ref{fig:fraction} on the New York dataset. We can see that it is beneficial to learn an adaptive mapping during the knowledge transfer even when limited aligned examples are available to train the translator. TrNews improves relative 0.82\%, 0.77\%, 0.67\%, 0.64\% in terms of HR@10 performance over CST by varying among $\{90\%, 70\%, 50\%, 30\%\}$ respectively. So we think that the more aligned examples the translator has, the more benefits it achieves.

\noindent
{\bf Impact of sharing word embeddings} We investigate the benefits of sharing word embeddings between source and target domains. There is a word embedding matrix for each of the domains and we share the columns if the corresponding words occur in both domains. Take the New York dataset as an example, the size of the intersection of their word vocabularies is 11,291 while the union is 50,263. From the results in Table~\ref{tb:sharing-word-embedding} we can see that it is beneficial to share the word embeddings even when only 22.5\% words are intersected between them.

\begin{table}
\centering
\resizebox{.48\textwidth}{!}{
\begin{tabular}{|c | c| c| c| c| c| c|}
\hline
Sharing?   & HR@5    & HR@10   & NDCG@5  & NDCG@10 & MRR     & AUC \\
\hline
No      & 81.31   & 94.69  &  59.43  & 63.72  & 54.37  &  97.16 \\
\hline
Yes    & {\bf 82.60} & {\bf 95.15} & {\bf 60.78} & {\bf 64.83} & {\bf 55.70} & {\bf 97.28} \\
\hline
\end{tabular}
}
\caption{Impact of sharing word embeddings between source and target domains.}
\label{tb:sharing-word-embedding}
\end{table}

\begin{table}
\centering
\resizebox{.48\textwidth}{!}{
\begin{tabular}{|c | c| c| c| c| c| c|}
\hline
Strategy   & HR@5    & HR@10   & NDCG@5  & NDCG@10 & MRR     & AUC \\
\hline
Sepa.      & 82.36 & 94.86 & 60.62  & 64.65 & 55.51 & 97.28 \\
\hline
Alter.    & {\bf 82.60} & {\bf 95.15} & {\bf 60.78} & {\bf 64.83} & {\bf 55.70} & {\bf 97.28} \\
\hline
\end{tabular}
}
\caption{Training TrNews with alternating (Alter.) vs separating (Sepa.) strategies.}
\label{tb:training-strategy}
\end{table}

\noindent
{\bf Impact of alternating training} We adopt an alternating training strategy between training the two (source \& target) networks and training the translator in our experiments. In this section, we compare this alternating strategy with the separating strategy which firstly trains the two networks and then trains the translator after completing the training of the two networks. That is, the training pairs of user representations for the translator are not generated on-the-fly during the training of source and target networks but generated only once after finishing their training. From the results in Table~\ref{tb:training-strategy}, we see that the alternating strategy works slightly better. This is probably because the aligned data between domains is limited and the alternating strategy increases the size of training pairs.

\noindent
{\bf Impact of two-stage learning} We adopt a two-stage model learning between training the two (source \& target) networks and training the translator in our experiments. In this section, we compare this two-stage learning with an end-to-end learning which jointly trains the two networks and the translator. That is, the parameters of the translator depend on the word embedding matrix and on parameters of the user encoder. From the results in Table~\ref{tb:training-two-stage}, we see that the two-stage learning works slightly better. This is probably because the aligned data between domains is too limited to reliably update the parameters which do not belong to the parameters of the translator.

\begin{figure}
\centering
\begin{subfigure}{.235\textwidth}
    \includegraphics[height=1.4in,width=4.2cm]{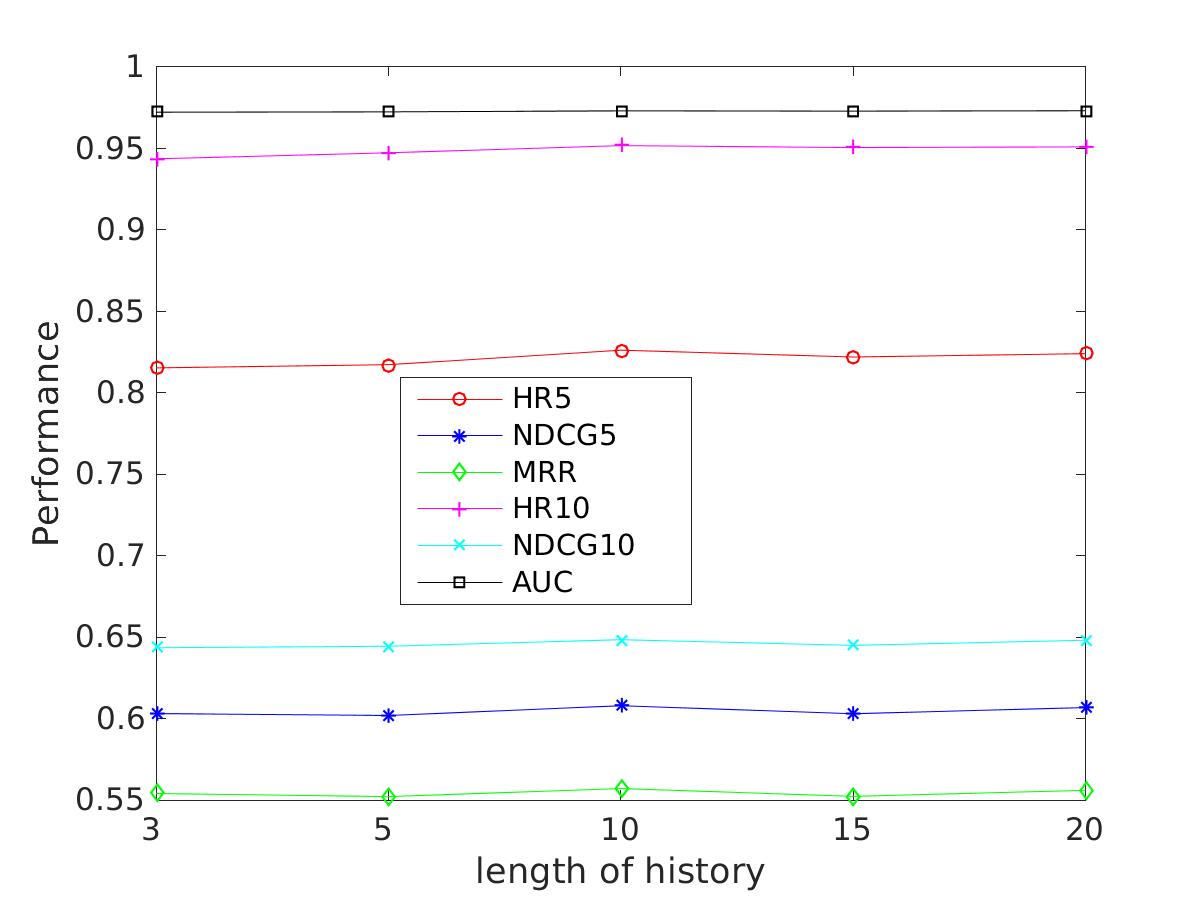}
\caption{History length.}
\label{fig:history}
\end{subfigure}
\begin{subfigure}{.235\textwidth}
      \includegraphics[height=1.4in,width=4.2cm]{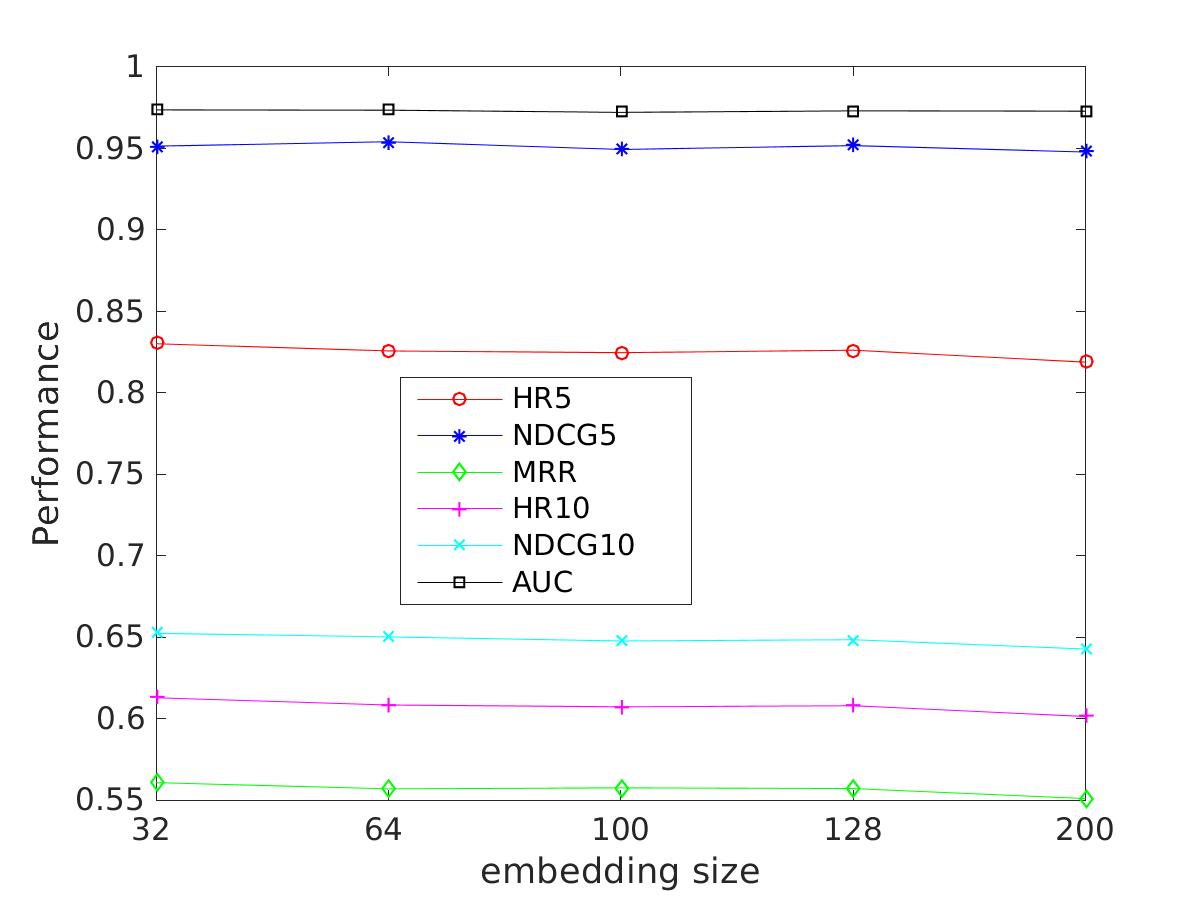}
\caption{Embedding size.}
\label{fig:embedding}
\end{subfigure}
\caption{Impact of the history length (left) and embedding size (right).  }
\label{fig:length-size}
\end{figure}

\noindent
{\bf Impact of the length of the history} Since we generate the training examples by sliding over the whole reading history for each user, the length of reading history is a key parameter to influence the performance of TrNews. We investigate how the length of the history affects the performance by varying it with $\{3, 5, 10, 15, 20\}$. The results on the New York dataset are shown in Figure~\ref{fig:history}. We can observe that increasing the size of the sliding window is sometimes harmful to the performance, and TrNews achieves good results for length 10. This is probably because of the characteristics of news freshness and of the dynamics of user interests. That is, the latest history matters more in general. Also, increasing the length of the input makes the training time increase rapidly, which are 58, 83, 143, 174, and 215 seconds when varying with $\{3, 5, 10, 15, 20\}$ respectively.

\begin{figure*}
\centering
\begin{subfigure}{.32\textwidth}
    \includegraphics[height=1.2in,width=5.3cm]{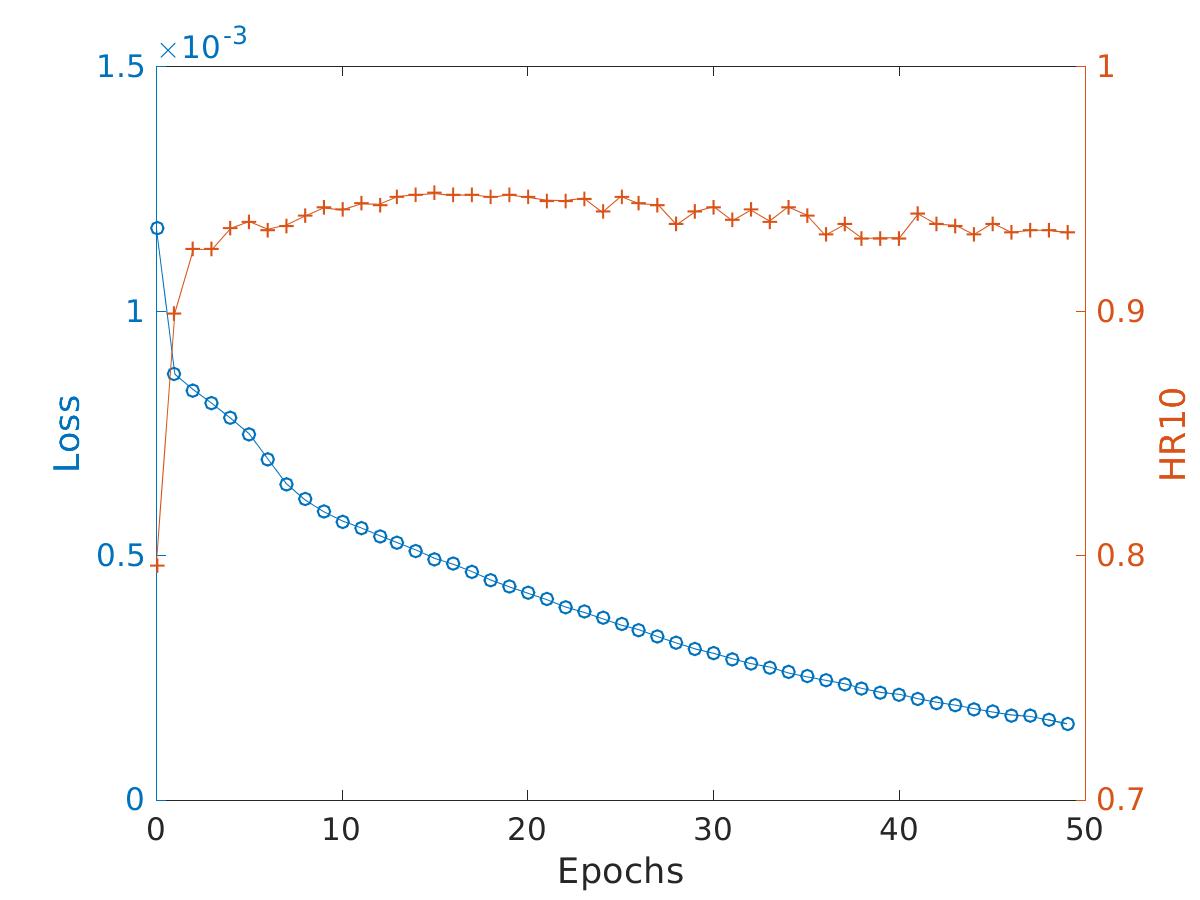}
    \caption{HR@10.}
    \label{fig:opt-hr}
\end{subfigure}
\begin{subfigure}{.32\textwidth}
    \includegraphics[height=1.2in,width=5.3cm]{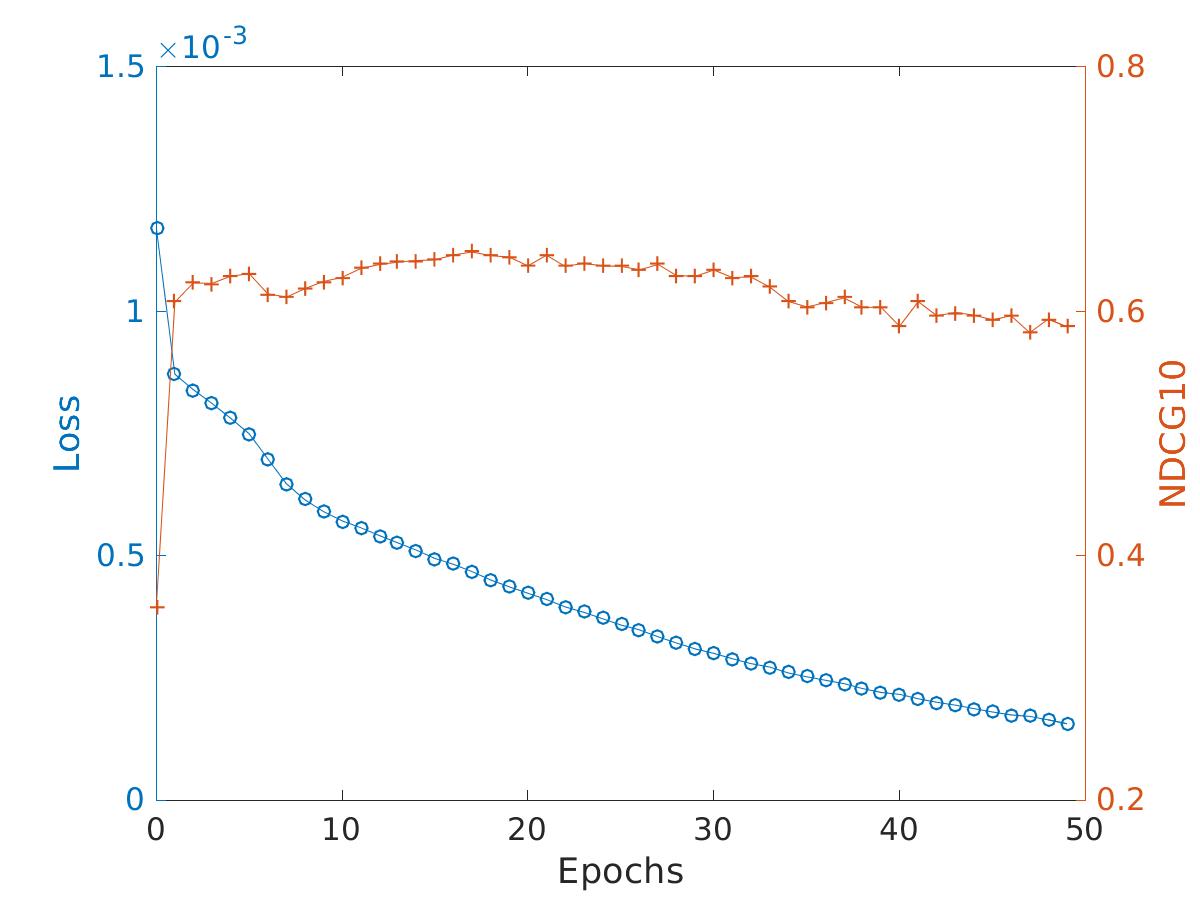}
    \caption{NDCG@10.}
    \label{fig:opt-ndcg}
\end{subfigure}
\begin{subfigure}{.32\textwidth}
    \includegraphics[height=1.2in,width=5.3cm]{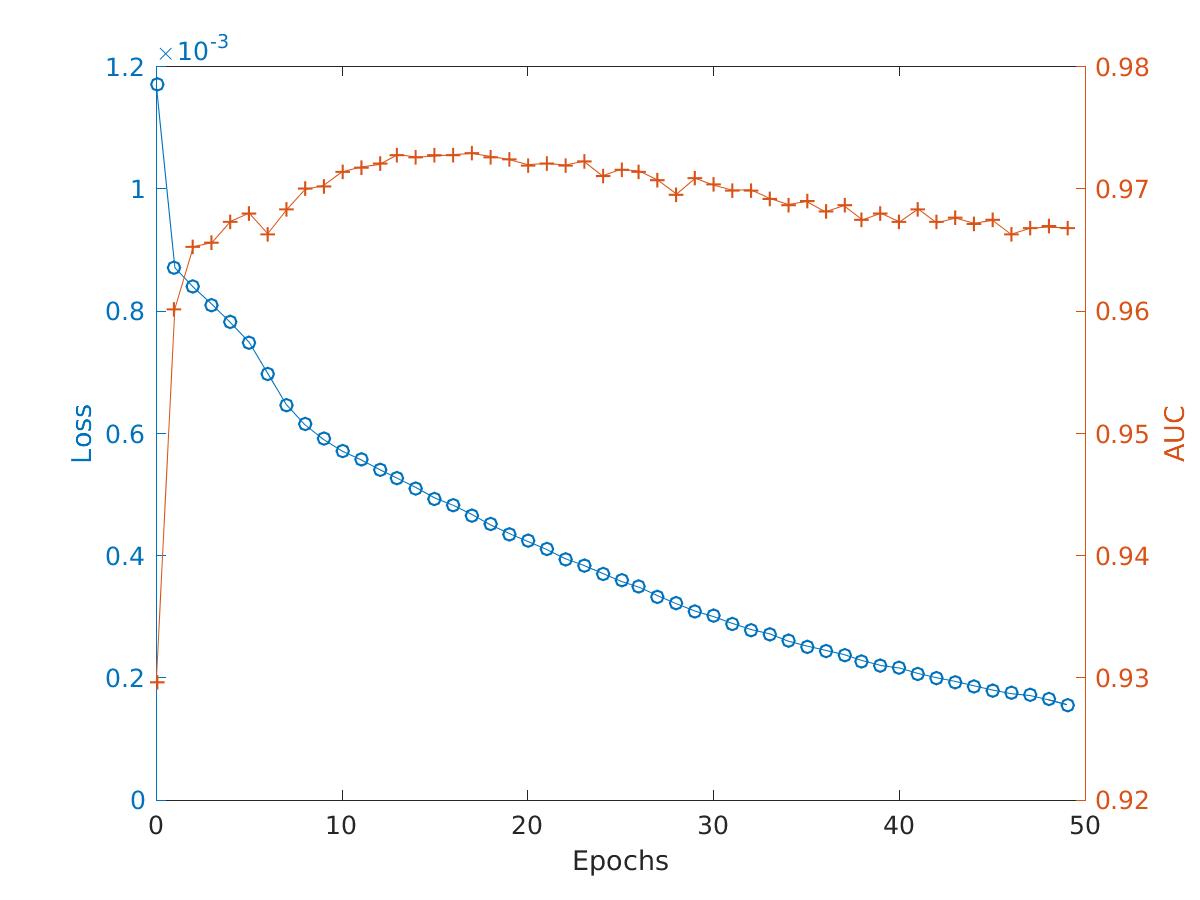}
    \caption{AUC.}
    \label{fig:opt-auc}
\end{subfigure}%
\caption{Performance (right Y-axis in red cross) and loss (left Y-axis in blue circle) varying with training iterations. }
\label{fig:optimizaton}
\end{figure*}

\noindent
{\bf Impact of the embedding size} In this section, we evaluate how different choices of some key hyperparameter affect the performance of TrNews. Except for the parameter being analyzed, all other parameters remain the same. Since we compute the news and user representations using the content of words, the size of word embedding is a key parameter to influence representations of words, uses, and news articles, and hence the performance of TrNews. We investigate how embedding size affects the performance by varying it with $\{32, 64, 100, 128, 200\}$. The results on the New York dataset are shown in Figure~\ref{fig:embedding}.  We can observe that increasing the embedding size is generally not harmful to the performance until 200, and TrNews achieves good results for embedding size 128. Changing it to 200 harms the performance a little bit since the model complexity also increases.

\noindent
{\bf Optimization performance and loss} We show the optimization performance and loss over iterations on the New York dataset in Figure~\ref{fig:optimizaton}.  We can see that with more iterations, the training losses gradually decrease and the recommendation performance is improved accordingly. The most effective updates are occurred in the first 15 iterations, and performance gradually improves until 30 iterations. With more iterations, TrNews is relatively stable. For the training time, TrNews spends 143 seconds per iteration. As a reference, it is 134s for DIN and 139s for TCB, which indicates that the training cost of TrNews is efficient by comparing with baselines. Furthermore, the test time is 150s. The experimental environment is Tensorflow 1.5.0 with Python 3.6 conducted on Linux CentOS 7 where The GPU is Nvidia TITAN Xp based on CUDA V7.0.27.

\noindent
{\bf Examining user profiles} One advantage of TrNews is that it can explain which article in a user's history matters the most for a candidate article by using attention weights in the user encoder module. Table~\ref{tb:case-study-1} shows an example of interactions between some user's history articles No. 0-9 and a candidate article No. 10, i.e., the user reads the candidate article after read these ten historical articles. We can see that the latest three articles matter the most since the user interests may remain the same during a short period. The oldest two articles, however, also have some impact on the candidate article, reflecting that the user interests may mix with a long-term characteristic. TrNews can capture these subtle short- and long-term user interests. 

\begin{table}
\centering
\resizebox{.48\textwidth}{!}{
\begin{tabular}{|c | c| c| c| c| c| c|}
\hline
Training   & HR@5    & HR@10   & NDCG@5  & NDCG@10 & MRR     & AUC \\
\hline
end-to-end      & 81.85 & 94.63 & 60.58 & 64.74 & 55.68 & 97.03 \\
\hline
two-stage    & {\bf 82.60} & {\bf 95.15} & {\bf 60.78} & {\bf 64.83} & {\bf 55.70} & {\bf 97.28} \\
\hline
\end{tabular}
}
\caption{Training TrNews in two-stage vs end-to-end.}
\label{tb:training-two-stage}
\end{table}

\begin{table}\small
\centering
\resizebox{0.5\textwidth}{!}{
\begin{tabular}{|c | c | c | }
\hline
No.    & News title &  \makecell{Attn. \\ weight}  \\
\hline
0 & hillary clinton makes a low-key return to washington  & 0.04 \\
\hline
1  & the hidden message in obama's `farewell' speech  & 0.12* \\
\hline
2  & here's why sasha obama skipped the farewell address  & 0.00 \\
\hline
3  & \makecell{donald trump's `prostitute scandal' was filmed by cameras \\  and recorded with microphones hidden behind the walls}  & 0.00 \\
\hline
4  & white house official explains sasha obama's absence at father's farewell speech   & 0.00 \\
\hline
5  & irish bookie puts odds on trump's administration, inauguration and impeachment   & 0.00 \\
\hline
6  & heads are finally beginning to roll at the clinton foundation   & 0.00 \\
\hline
7  & \makecell{donald trump's incoming administration considering \\ white house without press corps}   & {\bf 0.76} \\
\hline
8  & donald trump says merkel made `big mistake' on migrants   & 0.05 \\
\hline
9  & controversial clinton global initiative closing its doors for good   & 0.00 \\
\hline
10 & \makecell{ army chief gen. bipin rawat talks about equal responsibility \\  for women in the frontlines. we couldn't agree more}  & N/A \\
\hline
\end{tabular}
}
\caption{Example I: Some articles matter more while some are negligible. (No. 10 is the candidate news)}
\label{tb:case-study-1}
\end{table}

\section{Conclusion}

We investigate the cross-domain news recommendation via transfer learning. The experiments on real-word datasets demonstrate the necessity of tackling heterogeneity of user interests and word distributions across domains. Our TrNews model and its translator component are effective to transfer knowledge from the source network to the target network. We also shows that it is beneficial to learn a mapping from the source domain to the target domain even when only a small amount of aligned examples are available. In future works, we will focus on preserving the privacy of the source domain when we transfer its knowledge.

\section*{Acknowledgement}
We thank Dr. Yu Zhang for insightful discussion. The work is partly supported by the National Key Research and Development Program of China under Grant No.2018AAA0101100 and a collaboration project with DiDi: 19201730.

\bibliography{eacl2021}
\bibliographystyle{acl_natbib}

\end{document}